\documentclass[nobibnotes,nofootinbib]{revtex4}

\usepackage{amssymb,amsmath}
\usepackage{epsfig}

\begin{document}
\title{Saturation QCD predictions with heavy quarks at HERA}
\author{G. Soyez\footnote{on leave from the PTF group of the
    University of Li\`ege.}}  \email{g.soyez@ulg.ac.be}
\affiliation{Physics Department, Brookhaven National Laboratory,
  Upton, NY 11973, USA}

\begin{abstract}
  The measurement of the proton structure function at HERA is often
  seen as a hint for the observation of saturation in high-energy QCD
  {\em e.g.} through the observation of geometric scaling.
  Accordingly, the dipole picture provides a powerful framework in
  which the QCD-based saturation models can be confronted to the data.
  In this paper, we give a parametrisation of proton structure
  function which is directly constrained by the dynamics of QCD in its
  high-energy limit and fully includes the heavy quark effects. We
  obtain a good agreement with the available data. Furthermore, to the
  contrary of various models in the literature, we do not observe a
  significant decrease of the saturation momentum due to the heavy
  quark inclusion.
\end{abstract}

\newcommand{\abar}{\bar\alpha}

\maketitle

\section{Introduction}\label{sec:intro}

The study of the high-energy limit of perturbative QCD has lead to
strong predictions for the scattering amplitudes. One of the most
important result is the property of {\em geometric scaling} which is a
consequence of saturation \cite{iil,mt,t,mp} which extends into the
dilute regime where the amplitude is far from the unitarity limit. We
shall recall later in this paper how this general property can be
proven in perturbative QCD from the Balitsky-Kovchegov (BK) equation
\cite{bk} or from the Colour Glass Condensate (CGC) formalism \cite{cgc}.

At small $x$, the confrontation of those predictions with the
experimental measurements of the proton structure function at HERA can
be achieved within the framework of the dipole model. In the dipole
frame, the virtual photon fluctuates into a $q\bar q$ pair of flavour
$f$ and size $r$
which then interacts with the proton:
\begin{equation}\label{eq:sigma}
  \sigma_{L,T}^{\gamma^*p}(Q^2,x) = \sum_f \int d^2r\,\int_0^1 dz\,
  |\Psi_{L,T}^{(f)}(r,z;Q^2)|^2\,\sigma_{\text{dip}}(r,x).
\end{equation}
The photon wavefunction can safely be computed in QED and is found to be
\begin{eqnarray*}
|\Psi_{L,T}^{(f)}(r,z;Q^2)|^2
   & = & e_f^2 \frac{\alpha_e N_c}{2\pi^2}\, 4 Q^2 z^2(1-z)^2 K_0^2(r\bar Q_f),\\
|\Psi_{L,T}^{(f)}(r,z;Q^2)|^2
   & = & e_f^2 \frac{\alpha_e N_c}{2\pi^2}\left\{
           [z^2+(1-z)^2]\bar Q_f^2 K_1^2(r\bar Q_f)
          + m_f^2K_0^2(r\bar Q_f) \right\},
\end{eqnarray*}
where $\bar Q_f^2 = z(1-z)Q^2+m_f^2$. The proton structure function,
obtained through
\[
F_2(Q^2,x) = \frac{Q^2}{4\pi^2\alpha_e}
\left[\sigma_L^{\gamma^*p}(Q^2,x)+\sigma_T^{\gamma^*p}(Q^2,x)\right],
\]
is thus expressed in terms of the dipole-target cross-section
$\sigma_{\text{dip}}=2\pi R_p^2 T(r,x)$ where $T$ is the dipole-target
scattering amplitude as entering the QCD evolution equations. $R_p$ is
often referred to as the radius of the proton and is to be taken as a
free normalisation parameter of our model. We are therefore left with
the parametrisation of $T(r;Y)$, with $Y=\log(1/x)$ called the
rapidity.

This kind of approach is not new in itself. Different approaches to
parametrise the dipole-proton scattering amplitude has already been
proven successful. One can cite {\em e.g.} the pioneering work of
Golec-Biernat and Wusthoff \cite{gbw} which can be improved by adding
the collinear DGLAP effects \cite{bgk,dfgs}. Those approaches, directly
based on the gluon distribution function $xg(x,Q^2)$, give good
description of the data, even including the contributions from heavy
quarks \cite{sapeta}. From a different point of view, some other
successful approaches \cite{fs} use Regge parametrisations for the
dipole-proton scattering amplitude.

Beside those various approaches, perturbative QCD provides definite
predictions for $T$, including its approach to saturation. This is
perfectly suited for this kind of problem. Indeed, the factorisation
formula (\ref{eq:sigma}) underlying the dipole picture is valid only
in the high-energy (small-$x$) limit. Hence, using prediction from QCD
at high energy to parametrise the dipole-proton scattering amplitude
appears as a natural way to proceed. Those predictions that we shall
recall in is paper have been successfully gathered into a
parametrisation for $\sigma_{\text{dip}}(r,x)$ and tested against the
HERA data \cite{iim}, so far including the contributions from light
quarks only. All those approach suggest a {\em saturation scale}
$Q_s$, the energy-dependent momentum scale below which the amplitude
is saturated, of order 1 GeV$^2$ for $x\sim 10^{-4}-10^{-5}$ at HERA.

However, it is rather well-known that, once including the heavy quarks
(mainly the charm), all approaches observe a decrease of the
saturation scale. Given those {\em a priori} important effects of the
heavy quarks on the saturation, it appears important to reconsider the
predictions of the QCD at high energy to include those contributions.
In this paper, we will obtain two results: first that it is possible
to accommodate the predictions from \cite{iim} with heavy-quark
contributions and, second, that this does not lead to a decrease of
the saturation scale.

In the next section, we shall recall how to build the dipole-target
scattering amplitude $T$ from the equations of QCD at saturation. We
also discuss the idea allowing for the inclusion of heavy quarks.  We
shall then present the fit to the $F_2$ HERA data in itself, including
data selection and parameter adjustment. As a conclusion, we shall
finally discuss our results w.r.t. other models as well as with
predictions from NLO BFKL \cite{bfkl,nlo_bfkl,css_s}.

\section{QCD predictions for the amplitude and heavy quarks}\label{sec:qcd}

Since our approach is mainly based on the QCD fit introduced by Iancu,
Itakura and Munier (IIM) \cite{iim}, we start by a presentation of
that model. The formula we use to parametrise the dipole-proton
scattering amplitude is obtained from our knowledge of the solutions
of the BK equation which captures the main ingredients of the
high-energy physics with saturation effects. The exact solution to
that equation is not known but its asymptotic behaviours, for large
and small dipole sizes has been studied in details.

In the last years, it has been shown \cite{mp} that the BK equation
lies in the same class of universality than the
Fisher--Kolmogorov-Petrovsky-Piscunov (F-KPP) equation. The latter has
been extensively studied in statistical physics over the past seventy
years and it is well-known that its solutions can be written in terms
of {\em travelling waves}. In the language of the QCD variables we
have used so far this means that if one looks at the rapidity
evolution of the amplitude $T(r;Y)$ (seen here as a function of $r$),
the amplitude ``front'' moves towards smaller values of $r$ without
changing its shape. The ``position'' of the wavefront is then
naturally associated with $Q_s^{-1}$, the inverse of the saturation
scale. At asymptotic rapidities, the amplitude $T$, initially a
function of $r$ and $Y$ independently becomes a function of the single
variable $rQ_s(Y)$. This very important consequence of saturation is
the {\em geometric scaling} \cite{geomscaling,geomdiffr,geomsyst}
property which physically means that the physics remains unchanged
when one moves parallel to the saturation line.

This travelling-wave analysis provides two fundamental pieces of
information: first, the saturation scale increases exponentially with
the rapidity\footnote{Within that formalism, it is also possible to
  compute the next two subdominant terms but those are beyond the
  scope of the present paper.}: $Q_s^2(Y) \propto \exp(\lambda Y)$.
Then, the amplitude is known in the small-$r$ region:
\begin{equation}\label{eq:Tsmallr}
  T(r;Y) \propto
  \exp\left[-\gamma_c(\rho-\rho_s)
            -\frac{(\rho-\rho_s)^2}{2\abar\chi''_c Y} \right]
\end{equation}
with $\rho = \log(4/r^2)$ and $\rho_s = \log(Q_s^2)$. The critical
slope $\gamma_c$ as well as the parameter $\lambda=\abar\chi'_c$ and
$\chi''_c$ are determined from the linear BFKL kernel only. This is an
important property: though the scattering amplitude fully satisfies
the unitarity constraints and is sensitive to saturation effects, the
parameters which describes it do not depend upon the details of how
saturation is encoded. We will discuss the value for those parameters
later in this section. The fact that they do not depend on the details
of the saturation mechanism is another interesting feature of this
approach. As a related comment, one can also obtain \cite{iil,mt} the
result (\ref{eq:Tsmallr}) by looking at the BFKL equation with a
boundary condition at the saturation line: performing a saddle-point
approximation gives (\ref{eq:Tsmallr}) as well as the exponential
behaviour of the saturation momentum. Note finally that the Gaussian
part of the exponential in (\ref{eq:Tsmallr}) violates geometric
scaling as it introduces an explicit dependence in $Y$. This term
however becomes less and less important as rapidity increases. It
controls how geometric scaling is approached and allows to deduce that
geometric scaling is valid within a window $\rho-\rho_s \lesssim
\sqrt{2\abar \chi''_c Y}$. This is an important point that saturation
effects are relevant up to large scales above the saturation momentum
{\em i.e.} in the dilute domain.

The amplitude in the saturated domain is also obtained from the BK
equation \cite{BKsat} (it can also be obtained from the Colour Glass
Condensate formalism \cite{CGCsat}). Putting it together with
equation (\ref{eq:Tsmallr}) we reach the final expression for our
dipole-proton scattering amplitude:
\begin{equation}\label{eq:T}
  T(r;Y) = 
  \begin{cases}
    T_0 \exp\left[-\gamma_c(\rho-\rho_s)
                  -\frac{(\rho-\rho_s)^2}{2\kappa\lambda Y}  \right]
      & \text{if }rQ_s \le 2,\\
    1-\exp\left[-a(\rho-\rho_s-b)^2\right] & \text{if }rQ_s > 2,
  \end{cases}
\end{equation}
where the parameters $a$ and $b$ are fixed so as to ensure that $T$
and its derivative are continuous at $rQ_s=2$. Going from
(\ref{eq:Tsmallr}) to (\ref{eq:T}), we have used $\abar \chi''_c =
(\chi''_c/\chi'_c)\lambda Y = \kappa\lambda Y$ with
$\kappa=\chi''_c/\chi'_c$.

Let us now discuss the parameters in this QCD-saturation-based
model. First, the saturation scale can be written
\begin{equation}\label{eq:Qs}
Q_s^2(Y) = \left(\frac{x_0}{x}\right)^\lambda\text{ GeV}^2.
\end{equation}
This leaves two free parameters: $x_0$ which is related to the value
of the saturation scale at zero rapidity and corresponds to the value
of $x$ at which $Q_s=1$ GeV, and $\lambda$ which controls the rapidity
evolution of the saturation momentum. While leading-order (LO) BFKL
predicts $\lambda=\abar\chi'_c\approx 0.9$, an analysis of the
next-to-leading order (NLO) BFKL \cite{t} gives $\lambda\sim 0.3$, a
value which is in much better agreement with the phenomenological
analysis. 

Concerning the parameters in the amplitude itself, we shall fix the
matching amplitude $T_0$. As in \cite{iim}, a default value of 0.7
gives good results and variations around that value leads only to
small differences, so we will adopt $T_0=0.7$. The value of $\kappa$
will be set from the LO BFKL kernel which gives $\kappa\approx
9.9$. Note that the NLO BFKL predictions, though a little bit smaller,
remain of the same order. 

The value of the critical slope $\gamma_c$ is a fundamental issue of
this paper. In the original work \cite{iim}, where only the light
quarks were considered, the authors has fixed it to the value obtained
from the LO BFKL kernel ($\gamma_c\approx 0.6275$). However, as we
shall see in section \ref{sec:fit}, when including the heavy quarks,
keeping that value leads to a dramatic decrease of the saturation
momentum together as well as to a poor $\chi^2$ for the fit. A key
issue of the present work is to show that, allowing that parameter to
vary, we recover a similar saturation scale and a good fit. In
addition, we shall see that the value for $\gamma_c$ coming out of the
fit is rather close to what we expect from NLO BFKL ($\gamma_c \gtrsim
0.7$).

The last parameter entering the dipole-proton cross-section
$\sigma_{\text{dip}}(r,x)$ is the radius of the proton $R_p$ which
fixes the normalisation w.r.t. the dipole-proton amplitude $T(r,x)$.
We are thus left with 4 parameters: $R_p$, $x_0$ and $\lambda$ which
were already present in \cite{iim}, and the new one: $\gamma_c$.  In
the massless case, it was found, for $T_0=0.7$, $\lambda=0.253$,
$x_0=2.67\,10^{-5}$ and $R_p=0.641$ fm for $\gamma_c$ fixed to 0.6275.

Last but not least, we have to specify our heavy-quark prescriptions.
While in \cite{iim} the sum in \eqref{eq:sigma} was restricted to the
three light flavours only, we now want to consider the effect of
massive charm and bottom quarks. Those are thus introduced in the sum
over all flavours. The quark masses, entering the photon wavefunction
are fixed to $m_{u,d,s} = 140$ MeV, $m_c = 1.4$ GeV and $m_b = 4.5$
GeV and we have used the modified Bjorken variable $x(1+4
m_f^2/Q^2)$ in the contribution of the heavy quarks. Note that the
contribution of the charm and bottom quark to \eqref{eq:sigma}
directly give the charm and bottom structure functions.

\section{Fit to the HERA data}\label{sec:fit}

Now that we have fully described the saturation-based model we are
using to describe the DIS structure function, we will test its
validity by fitting its free parameters to the experimental
measurements of the proton structure function $F_2$. In this section
we present the details of the fit and discuss the results.

We first have to specify which dataset we are working with. Following
the recent analysis, we shall use the last HERA data {\em i.e.} the
last ZEUS\cite{zeus} and H1\cite{h1} measurements of $F_2$. We include
a 5\% renormalisation uncertainty on the H1 data to account for a
normalisation discrepancy between ZEUS and H1. Note that the analysis
in \cite{iim} only takes into account the ZEUS data. We shall come
back on this point later, when we turn to the discussion of our
results.

Our approach is focused on the high-energy behaviour of DIS: the
dipole-model factorisation \ref{eq:sigma} is only valid at
sufficiently small $x$ and, accordingly, the dipole-proton amplitude
is build from the high-energy QCD equations. Hence, we shall limit
ourselves to $x\le 0.01$, the usual cut in those approaches.
Furthermore, our approach, based on the QCD equations describing
saturation, does not takes into account the DGLAP corrections beyond
the double-logarithmic approximation. We shall restrict our analysis
to $Q^2\le 150$ GeV$^2$. Note that the former analysis with massless
quarks \cite{iim} is more conservative and uses $Q^2\le 45$
GeV$^2$. However, in order to check the validity of the results
discussed hereafter, we have tested both cuts and observed that the
$\chi^2$ and the parameters were not significantly changing when
increasing the $Q^2$ cut. 

We have fitted the free parameters of the model discussed in Section
\ref{sec:qcd} to the 281 data contained in our dataset. In order to
grasp the effect of the correct treatment of the heavy-quark masses,
we performed the fit with and without including the charm and bottom
contribution to $F_2$. In addition, for both cases we give the result
of the fit for the critical slope $\gamma_c$ fixed to its LO value or
considered as a free parameter. The resulting parameters and $\chi^2$
are presented in table \ref{tab:res} where we have also added the
initial parameters from \cite{iim} (with ZEUS data only and $Q^2\le
45$ GeV$^2$) for better comparison. 

\begin{table}
\begin{tabular}{|l|l||c|c|c|c||c|}
\hline
  &  & $\gamma_c$ & $\lambda$ & $x_0$ ($10^{-4}$) & $R_p$ (GeV$^{-1}$) &  $\chi^2/$n.o.p. \\
\hline\hline
light quarks only & \cite{iim} (ZEUS only) &       0.6275        
                                  &       0.253         
                                  &       0.267          
                                  &       3.25          &  -   \\
\cline{2-7}
         & $\gamma_c$ fixed       &       0.6275        
                                  & 0.2574 $\pm$ 0.0037 
                                  & 0.2750 $\pm$ 0.0240  
                                  & 3.241  $\pm$ 0.018  & 0.959 \\
\cline{2-7}
         & $\gamma_c$ free        & 0.6194 $\pm$ 0.0091 
                                  & 0.2545 $\pm$ 0.0051 
                                  & 0.2131 $\pm$ 0.0651  
                                  & 3.277  $\pm$ 0.044  & 0.956 \\
\hline\hline
light+heavy quarks  & $\gamma_c$ fixed       &       0.6275 
                                  & 0.1800 $\pm$ 0.0026
                                  & 0.0028 $\pm$ 0.0003
                                  & 3.819  $\pm$ 0.017  &  1.116\\
\cline{2-7}
         & $\gamma_c$ free        & $\mathbf{0.7376 \pm 0.0094}$
                                  & $\mathbf{0.2197 \pm 0.0042}$
                                  & $\mathbf{0.1632 \pm 0.0471}$
                                  & $\mathbf{3.344  \pm 0.041 }$  &
                                  $\mathbf{0.900}$ \\
\hline
\end{tabular}
\caption{The table gives the parameters and $\chi^2$ per point
  obtained from the fit. The results are shown with or without the
  heavy quark contribution and with $\gamma_c$ free or fixed. The last
  line of this table is the main result of this paper.}\label{tab:res}
\end{table}

\begin{figure}
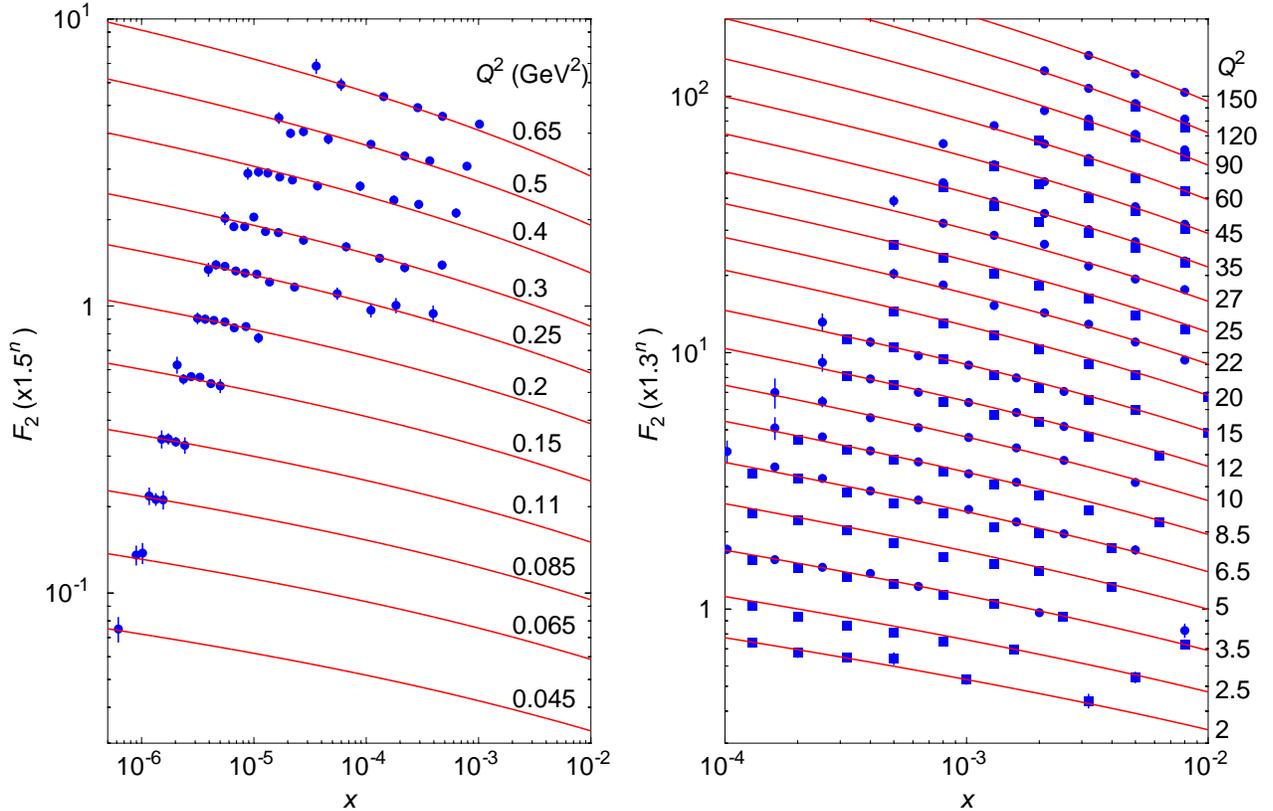

\centerline{
\includegraphics[scale=0.8]{f2p_lowQ2.ps}
\includegraphics[scale=0.8]{f2p_highQ2.ps}
}
\caption{Result of our fit to the proton structure function. The left
  plot shows the low $Q^2$ bins ($Q^2<1$ GeV$^2$) while larger $Q^2$
  are given on the right plot. The $Q^2$ value corresponding to each
  curve is given (in GeV$^2$). For clarity, successive curves, from
  bottom to top, have been rescaled by powers of 1.5.}\label{fig:f2p}
\end{figure}

Those results deserve some comments:
\begin{itemize}
\item Concerning the re-analysis of the fit without heavy quarks
  (second line of table \ref{tab:res}), we see that with the addition
  of the H1 data and the extension of the $Q^2$ domain, the parameters
  remain similar and the $\chi^2$ is good.
\item If one allows $\gamma_c$ to freely vary in the massless case
  (third line of table \ref{tab:res}), again, the fit naturally
  converges to a minimum which is close to the initial one without
  improving significantly the $\chi^2$. The LO choice
  $\gamma_c=0.6275$ is even compatible with the error bars of what we
  obtain when we fit it.
\item Once the heavy quarks are taken into account, the situation
  changes drastically. If the critical slope is fixed to its LO value,
  the situation becomes dramatic. Indeed, not only the quality of the
  fit is getting worse, but also the saturation scale is going down by
  two orders of magnitude (the exponent $\lambda$ is also decreasing
  significantly).
\item If as anticipated in Section \ref{sec:qcd} we allow the critical
  slope $\gamma_c$ to vary, the fit (last line of table \ref{tab:res})
  converges back to a good description (even a better $\chi^2$ than
  the corresponding massless fit). The parameters describing the
  saturation scale are also rather close to those obtained in the
  massless case.
\item The value of the critical slope we obtain from the fit seems
  much larger than the LO result used previously. However, if one
  extract that value from various renormalisation-group-improved NLO
  BFKL kernels \cite{css_s} one get a value of $\gamma_c$ slightly
  larger than 0.7. This is again in good agreement with the value to
  which the new fit naturally converges.
\item To test the dependence upon the fixed parameters of our model,
  we have performed various fits varying those parameters around their
  default value. We have thus changed the $Q^2$ cut from 150
  to 45 GeV$^2$, included only the heavy charm (without bottom),
  varied the masses of the heavy quarks and varied the matching point
  $T_0$. For all those variations, both the quality of the fit and the
  values of the parameters remained similar, which enforces the
  robustness of the present parametrisation.
\end{itemize}

\begin{figure}
\centerline{
\includegraphics[scale=0.66]{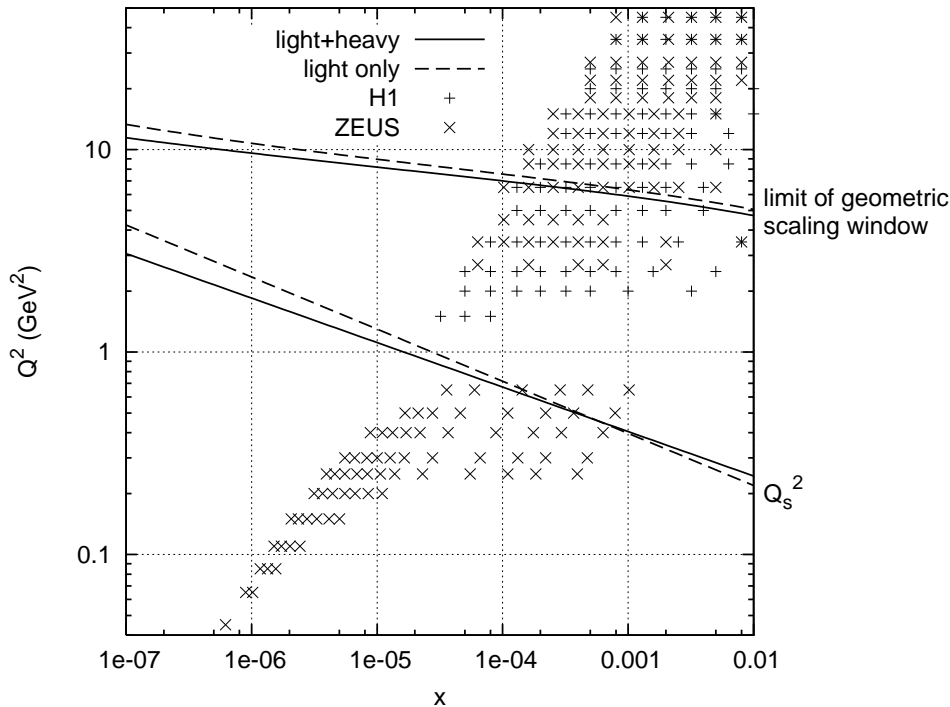}
}
\caption{For both the fit in this paper (light+heavy flavours, solid
  lines) and the fit of \cite{iim} (light flavours only, dashed
  lines), we plot two information: the saturation scale $Q_s^2(x)$ and
  the extension of the geometric scaling window $\log(Q^2/Q_s^2) \le
  \sqrt{2\kappa\lambda Y}$. We see that the inclusion of the heavy
  quarks does not lead to a strong decrease both those scales. The
  points on the plot represent the $(x,Q^2)$ position of the HERA
  measurements.}\label{fig:qs}
\end{figure}

The last line of table \ref{tab:res}, giving a description of the
saturation effects in DIS including the heavy quark effects, has to be
considered as the main result of this paper. The corresponding
description of the proton structure function is plotted in figure
\ref{fig:f2p}. 

One of the most interesting point of this parametrisation is that the
saturation scale obtained with heavy quarks included is very similar
to the one obtained with light quarks only. This is better seen in
figure \ref{fig:qs} where we have plot the saturation scale (lower
curves) as well as the limit of the geometric scaling window both with
and without heavy quarks contributions. One clearly see that the
addition of the heavy quark contribution only slightly reduces the
effect of saturation. This is an important result as previous models
including heavy quark effects all report a decrease of the saturation
momentum by (roughly) a factor of 2 once those heavy quarks
contributions are included. From figure \ref{fig:qs} it also appears
that a large number of data (all data from small $Q^2$ up to the limit
of the geometric scaling window) are sensitive to saturation effects.

\begin{figure}
\centerline{\includegraphics{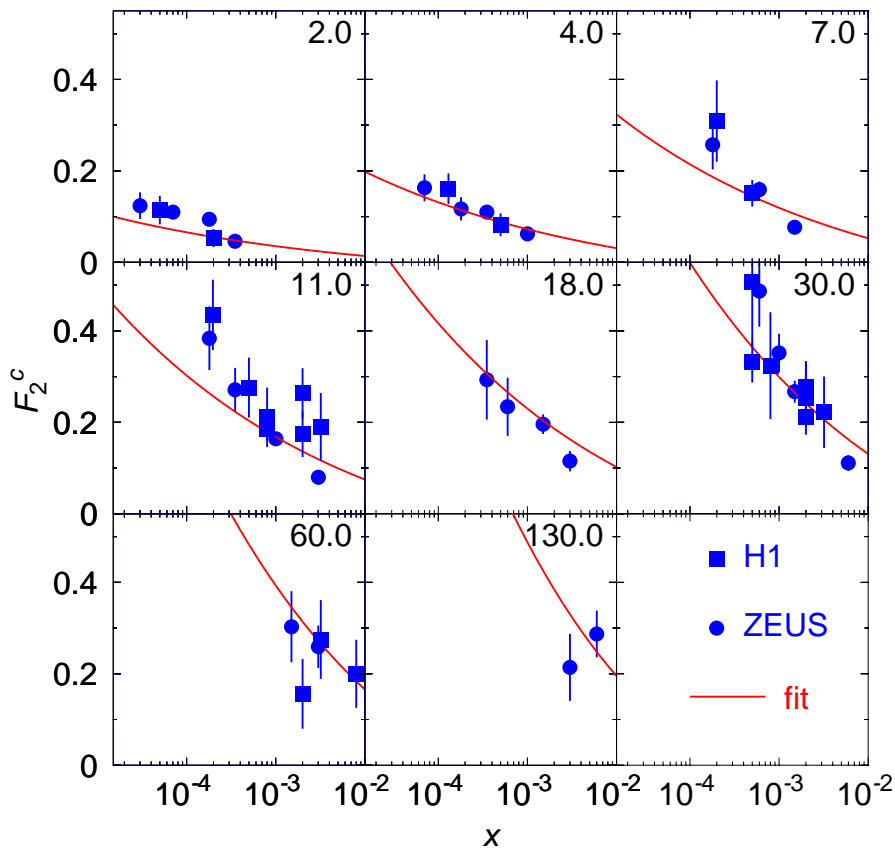}}
\caption{Predictions of our fit for the charm structure
  function.}\label{fig:f2c}
\end{figure}

\begin{figure}
\centerline{\includegraphics{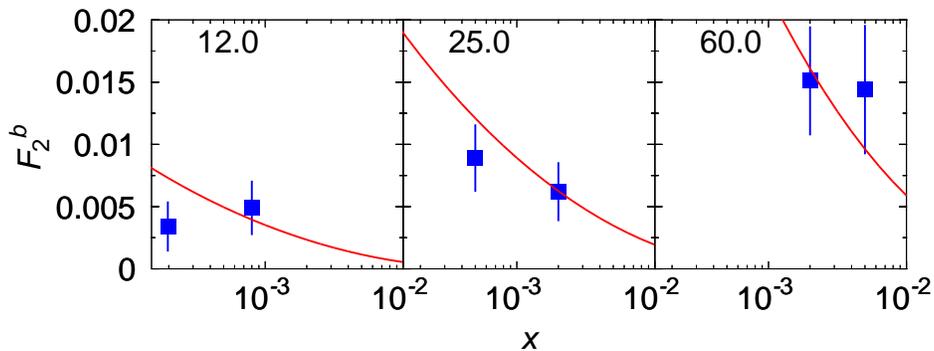}}
\caption{Predictions of our fit for the bottom structure
  function.}\label{fig:f2b}
\end{figure}

\begin{figure}
\centerline{\includegraphics{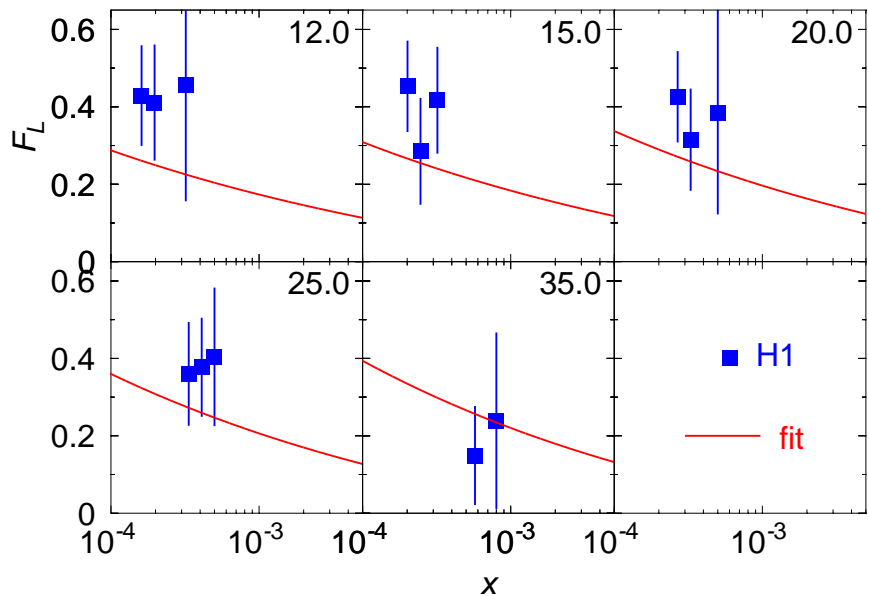}}
\caption{Predictions of our fit for the longitudinal structure
  function.}\label{fig:fl}
\end{figure}

Finally, we can compare the predictions of our parametrisation with
the HERA measurements \cite{h1c,zeusc} of the charm and bottom
structure functions. Those are naturally obtained from our formalism
by taking the charm or bottom contribution to the photon-proton
cross-section (\ref{eq:sigma}). The prediction for our model are
plotted in figures \ref{fig:f2c} and \ref{fig:f2b} for the charm and
bottom structure functions respectively. In both cases, we observe a
good agreement with the data. Similarly, by taking the contribution
coming from the longitudinal part of the wavefunction in
(\ref{eq:sigma}), we can obtain predictions for the longitudinal
structure function. Our result is shown in figure \ref{fig:fl}
together with the H1 measurements \cite{h1}. Again, the present
parametrisation gives a good description of the data.

\section{Discussion and conclusions}\label{sec:ccl}

In this paper, we have shown that it was possible to accommodate the
saturation model introduced by Iancu, Itakura and Munier \cite{iim} to
take into account the heavy quark contribution to the proton structure
function. The resulting fit provides a very good description of the
HERA measurements of $F_2^p$ and the predictions for the heavy-quark
structure functions as well as for the longitudinal structure function
are in good agreement with the existing data.

The present work, using the dipole picture formalism, has two
important features that distinguish it from previous studies in the
literature: firstly, following the idea in \cite{iim}, we have
directly use the predictions from high-energy QCD to parametrise the
dipole-proton amplitude. The saturation properties, {\em e.g.}
geometric scaling and its window of validity, are thus parametrised as
they are predicted from perturbative QCD. Secondly, the saturation
scale that result from our fit is not significantly reduced compared
to the saturation scale obtained with light quarks only. This
contrasts with previous studies in the literature which report a
decrease of the saturation momentum. In the present approach
saturation effects cannot be neglected. This new analysis gives an
additional argument in favour of saturation.

We also stress a recent analysis \cite{momentum} also including
heavy-quark effects and based on the high-energy QCD properties, where
the amplitude is parametrised in momentum space rather than in
coordinate space. Compared to that study, the parameters for the
saturation scale and geometric scaling window extension, both slightly
too small in the momentum-space parametrisation, are more reasonable
in the present approach. The analysis in momentum space was however
carried with fixed $\gamma_c$ and further studies are requested to
check whether a higher critical slope also improves the model.

A last comment concerns the relation between the parameters obtained
in our fit and the predictions from the NLO BFKL kernel with a
saturation boundary. Indeed, though a bit smaller, the exponent of the
saturation scale remains close to the predictions from the NLO BFKL
\cite{t}. In addition, we have seen that the critical slope $\gamma_c$
kept fixed to its leading-order value in the massless case, is no
longer in agreement with that value once the heavy quarks are taken
into account. The value we obtain in that case is rather in agreement
with the slope one would obtain from the
renormalisation-group-improved BFKL kernels at NLO. This finding is
another welcome outcome of our parametrisation.

The last point naturally suggests to directly compare the predictions
of NLO BFKL with saturation effects to the proton structure function.
However, to rigorously address that question, one probably also needs
to correctly introduce the running-coupling corrections to the present
formalism. Though they are not expected to lead to significant
modifications, they come with a few changes that we leave for future
work.

Finally, now that the coordinate-space dipole-proton scattering
amplitude is available as directly predicted from the QCD saturation
framework, applications to other observables are to be done. This
includes the diffractive structure function often considered as an
excellent candidate for the observation of saturation and
well-described in the dipole picture \cite{gbw,sapeta,fs,bpr,fs2}. In
addition, extending to non-zero momentum transfer (also predicted by
high-energy QCD \cite{bkfull}) or parametrising the impact-parameter
dependence, it can also be tested against the diffractive vector-meson
production and DVCS cross-sections \cite{kmw,fs3,vm}. Those
complementary analysis are also left for future work.

\begin{acknowledgments}
  I gratefully thank Edmond Iancu, Cyrille Marquet and Robi Peschanski
  for useful discussions and careful reading of the manuscript.  G.S.
  is funded by the National Funds for Scientific Research (FNRS,
  Belgium)
\end{acknowledgments}


\begin{thebibliography}{99}

\bibitem{iil}
E.~Iancu, K.~Itakura and L.~McLerran,
Nucl.\ Phys.\  A {\bf 708} (2002) 327 [arXiv:hep-ph/0203137].
%% CITATION = NUPHA,A708,327;%%

\bibitem{mt}
A.~H.~Mueller and D.~N.~Triantafyllopoulos,
Nucl.\ Phys.\  B {\bf 640} (2002) 331 [arXiv:hep-ph/0205167].
%% CITATION = NUPHA,B640,331;%%

\bibitem{t}
D.~N.~Triantafyllopoulos,
Nucl.\ Phys.\  B {\bf 648} (2003) 293 [arXiv:hep-ph/0209121].
%% CITATION = NUPHA,B648,293;%%

\bibitem{mp}
S.~Munier and R.~Peschanski,
Phys.\ Rev.\ Lett.\  {\bf 91}, 232001 (2003)
[arXiv:hep-ph/0309177];
%%CITATION = HEP-PH 0309177;%%
Phys.\ Rev.\ {\bf D69}, 034008 (2004)
[arXiv:hep-ph/0310357];
%%CITATION = HEP-PH 0309177;%%
Phys.\ Rev.\ {\bf D70}, 077503 (2004)
[arXiv:hep-ph/0310357].

\bibitem{bk}
I.~I.~Balitsky,
Nucl.\ Phys.\ {\bf B463}, 99 (1996) [arXiv:hep-ph/9509348];
%%CITATION = HEP-PH 9509348;%%
Y.~V. Kovchegov,
\newblock Phys. Rev. {\bf D60}, 034008 (1999) [arXiv:hep-ph/9901281];
%%CITATION = HEP-PH 9901281;%%
\newblock  Phys. Rev. {\bf D61}, 074018 (2000), [arXiv:hep-ph/9905214].
%%CITATION = HEP-PH 9905214;%%

\bibitem{cgc}
E.~Iancu and R.~Venugopalan,
arXiv:hep-ph/0303204.
%% CITATION = HEP-PH/0303204;%%

\bibitem{gbw}
K.~Golec-Biernat and M.~Wusthoff,
Phys.\ Rev.\  D {\bf 59} (1999) 014017 [arXiv:hep-ph/9807513];
%% CITATION = PHRVA,D59,014017;%%
Phys.\ Rev.\  D {\bf 60} (1999) 114023 [arXiv:hep-ph/9903358].
%% CITATION = PHRVA,D60,114023;%%

\bibitem{bgk}
J.~Bartels, K.~Golec-Biernat and H.~Kowalski,
Phys.\ Rev.\  D {\bf 66} (2002) 014001 [arXiv:hep-ph/0203258].
%% CITATION = PHRVA,D66,014001;%%

\bibitem{dfgs}
M.~McDermott, L.~Frankfurt, V.~Guzey and M.~Strikman,
Eur.\ Phys.\ J.\  C {\bf 16} (2000) 641 [arXiv:hep-ph/9912547].
%% CITATION = EPHJA,C16,641;%%

\bibitem{sapeta}
K.~Golec-Biernat and S.~Sapeta,
Phys.\ Rev.\  D {\bf 74} (2006) 054032 [arXiv:hep-ph/0607276].
%% CITATION = PHRVA,D74,054032;%%

\bibitem{fs}
J.~R.~Forshaw, G.~Kerley and G.~Shaw,
Phys.\ Rev.\  D {\bf 60} (1999) 074012 [arXiv:hep-ph/9903341].
%% CITATION = PHRVA,D60,074012;%%
J.~R.~Forshaw and G.~Shaw,
JHEP {\bf 0412} (2004) 052 [arXiv:hep-ph/0411337].
%% CITATION = JHEPA,0412,052;%%

\bibitem{iim}
E.~Iancu, K.~Itakura and S.~Munier,
Phys.\ Lett.\  B {\bf 590} (2004) 199 [arXiv:hep-ph/0310338].
%% CITATION = PHLTA,B590,199;%%

\bibitem{bfkl}
L. N. Lipatov, {\it Sov. J. Nucl. Phys.} {\bf 23}, (1976) 338;
E. A. Kuraev, L. N. Lipatov and V. S. Fadin,
{\it Sov. Phys. JETP} {\bf 45}, (1977) 199;
I. I. Balitsky and L. N. Lipatov,
{\it Sov. J. Nucl. Phys.} {\bf 28}, (1978) 822.

\bibitem{nlo_bfkl}
V.~S. Fadin and L.~N. Lipatov,
Phys. Lett. {\bf B429}, 127 (1998);
M.~Ciafaloni and G.~Camici,
Phys. Lett. {\bf B430}, 349 (1998).
%%CITATION = HEP-PH 9803389;%%

\bibitem{css_s}
G.~P. Salam,
JHEP {\bf 07}, 019 (1998);
%%CITATION = HEP-PH 9806482;%%
M.~Ciafaloni, D.~Colferai, and G.~P. Salam,
Phys. Rev. {\bf D60}, 114036 (1999).
%%CITATION = HEP-PH 9905566;%%

\bibitem{fkpp}
R.~A. Fisher,
\newblock Ann. Eugenics {\bf 7}, 355 (1937);
\newblock A.~Kolmogorov, I.~Petrovsky, and N.~Piscounov,
\newblock Moscou Univ. Bull. Math. {\bf A1}, 1 (1937).

\bibitem{geomscaling}
A.~M. Sta\'sto, K.~Golec-Biernat, and J.~Kwiecinski,
\newblock Phys. Rev. Lett. {\bf 86}, 596 (2001) [arXiv:hep-ph/0007192].
%%CITATION = HEP-PH 0007192;%%

\bibitem{geomdiffr}
C.~Marquet and L.~Schoeffel,
Phys.\ Lett.\  B {\bf 639} (2006) 471 [arXiv:hep-ph/0606079].
%% CITATION = PHLTA,B639,471;%%

\bibitem{geomsyst}
F.~Gelis, R.~Peschanski, G.~Soyez and L.~Schoeffel,
Phys.\ Lett.\ B {\bf 647} (2007) 376 [arXiv:hep-ph/0610435].
%% CITATION = HEP-PH/0610435;%%

\bibitem{BKsat}
E.~Levin and K.~Tuchin,
Nucl.\ Phys.\  B {\bf 573} (2000) 833 [arXiv:hep-ph/9908317].
%% CITATION = NUPHA,B573,833;%%

\bibitem{CGCsat}
E.~Iancu and L.~D.~McLerran,
Phys.\ Lett.\  B {\bf 510} (2001) 145 [arXiv:hep-ph/0103032];
%% CITATION = PHLTA,B510,145;%%
E.~Iancu and A.~H.~Mueller,
Nucl.\ Phys.\  A {\bf 730} (2004) 460 [arXiv:hep-ph/0308315].
%% CITATION = NUPHA,A730,460;%%

\bibitem{h1} 
C.~Adloff {\it et al.}  [H1 Collaboration],
Eur.\ Phys.\ J.\  C {\bf 21} (2001) 33 [arXiv:hep-ex/0012053].
%% CITATION = EPHJA,C21,33;%%

\bibitem{zeus} 
J.~Breitweg {\it et al.}  [ZEUS Collaboration],
Phys.\ Lett.\ B {\bf 487} (2000) 273 [arXiv:hep-ex/0006013];
%%CITATION = HEP-EX 0006013;%%
S.~Chekanov {\it et al.}  [ZEUS Collaboration],
Eur.\ Phys.\ J.\ C {\bf 21} (2001) 443 [arXiv:hep-ex/0105090].
%%CITATION = HEP-EX 0105090;%%

\bibitem{h1c} 
C.~Adloff {\it et al.}  [H1 Collaboration],
Phys.\ Lett.\  B {\bf 528} (2002) 199 [arXiv:hep-ex/0108039];
%% CITATION = PHLTA,B528,199;%%
A.~Aktas {\it et al.}  [H1 Collaboration],
Eur.\ Phys.\ J.\  C {\bf 45} (2006) 23 [arXiv:hep-ex/0507081].
%% CITATION = EPHJA,C45,23;%%

\bibitem{zeusc}
S.~Chekanov {\it et al.}  [ZEUS Collaboration],
Phys.\ Rev.\  D {\bf 69} (2004) 012004 [arXiv:hep-ex/0308068].
%% CITATION = PHRVA,D69,012004;%%

\bibitem{momentum}
J.~T.~de Santana Amaral, M.~B.~Gay Ducati, M.~A.~Betemps and G.~Soyez,
arXiv:hep-ph/0612091.
%% CITATION = HEP-PH/0612091;%%

\bibitem{bpr}
A.~Bialas, R.~Peschanski and C.~Royon,
Phys.\ Rev.\  D {\bf 57} (1998) 6899 [arXiv:hep-ph/9712216].
%% CITATION = PHRVA,D57,6899;%%

\bibitem{fs2}
J.~R.~Forshaw, G.~R.~Kerley and G.~Shaw,
Nucl.\ Phys.\  A {\bf 675} (2000) 80C [arXiv:hep-ph/9910251].
%% CITATION = NUPHA,A675,80C;%%
J.~R.~Forshaw, R.~Sandapen and G.~Shaw,
JHEP {\bf 0611} (2006) 025 [arXiv:hep-ph/0608161].
%% CITATION = JHEPA,0611,025;%%

\bibitem{bkfull}
C.~Marquet, R.~Peschanski and G.~Soyez,
Nucl.\ Phys.\  A {\bf 756} (2005) 399 [arXiv:hep-ph/0502020].
%% CITATION = NUPHA,A756,399;%%
C.~Marquet and G.~Soyez,
Nucl.\ Phys.\  A {\bf 760} (2005) 208 [arXiv:hep-ph/0504080].
%% CITATION = NUPHA,A760,208;%%

\bibitem{kmw}
H.~Kowalski and D.~Teaney,
Phys.\ Rev.\  D {\bf 68} (2003) 114005 [arXiv:hep-ph/0304189].
%% CITATION = PHRVA,D68,114005;%%
H.~Kowalski, L.~Motyka and G.~Watt,
Phys.\ Rev.\  D {\bf 74} (2006) 074016 [arXiv:hep-ph/0606272].
%% CITATION = PHRVA,D74,074016;%%

\bibitem{fs3}
J.~R.~Forshaw, R.~Sandapen and G.~Shaw,
JHEP {\bf 0611} (2006) 025 [arXiv:hep-ph/0608161];
%% CITATION = JHEPA,0611,025;%%
Phys.\ Rev.\  D {\bf 69} (2004) 094013 [arXiv:hep-ph/0312172].
%% CITATION = PHRVA,D69,094013;%%

\bibitem{vm}
C.~Marquet, R.~Peschanski and G.~Soyez, arXiv:hep-ph/0702171.
%% CITATION = HEP-PH/0702171;%%

\end{thebibliography}
\end{document}